\documentstyle[preprint,aps,epsf]{revtex}
\begin{document}
\draft
\preprint{}

\title{Hadronic Correlators and Condensate Fluctuations in QCD Vacuum } 
\author{Varun Sheel\footnote{Electronic address: varun@prl.ernet.in}, 
Hiranmaya Mishra\footnote{On leave from :
The Mehta Research Institute for Mathematics and Mathematical
Physics, 10 Kasturba Gandhi Marg, Allahabad - 211 022, India}
and Jitendra C. Parikh \footnote{Electronic address: parikh@prl.ernet.in} } 
\address{Theory Group, 
Physical Research Laboratory, Navrangpura, Ahmedabad 380 009, India}
\maketitle
\begin{abstract}
Phenomenological results of equal time, point to point spatial 
correlation functions of hadronic currents are used to deduce the structure
of the QCD vacuum. It is found that a model with only quark condensate is not
adequate to explain the observations. Inclusion of condensate fluctuations
(explicit four point structure in the vacuum) leads to excellent overall
agreement with the phenomenological curves and parameters in various
channels.

\end{abstract}
\vskip 0.5cm

The most interesting question in Quantum Chromodynamics (QCD) is concerned
with the nature of the vacuum state\cite{feynman}. It is well known that 
the vacuum is non-trivial and is composed of gluon and quark field 
condensates\cite{sur,surcor}. Much of the understanding of 
non-perturbative phenomena in QCD is expected to result from a proper
knowledge of the ground state of QCD.

In this note, we adopt a phenomenological approach to
determine the salient features of the QCD vacuum. We consider here
spacelike separated correlation functions of hadron currents in QCD vacuum
\cite{surcor}.
 More precisely, we use
phenomenological results of equal time, point to point spatial ground state
correlation functions of hadronic currents \cite{neglecor}
 to guide us towards a ``true''
structure of QCD vacuum.

As a first step, we employ the explicit construct for QCD vacuum with quark
and gluon condensates proposed by us\cite{ijmpe}. In this the trial ansatz for
the QCD vacuum was\cite{ijmpe}

\begin{equation}
|vac>=\exp{({B_F}^\dagger-B_F)}\exp{({B_G}^\dagger-B_G)}|0>
\label{vacp}
\end{equation}
where $|0>$ is the perturbative vacuum and ${B_F}^\dagger$ and
${B_G}^\dagger$ are the Bogoliubov operators corresponding to creation of
quark antiquark pairs and gluon pairs respectively. Such a construct
gives rise to the equal time propagator\cite{adler} as (with $x=|\vec x|$),

\begin{eqnarray}
S_{\alpha \beta}(\vec x) &  = & \left< vac | \frac{1}{2} \left[
 \psi_{\alpha}^{i}(\vec x),
             \bar \psi_{\beta}^{i}(0) \right] | vac \right>
\nonumber \\
 & = & \frac{1}{2} \frac{1}{(2 \pi)^3}
  \int e^{i\vec k .\vec x }d\vec k \left[ \sin 2 h(\vec k)
 - (\vec\gamma \cdot \hat k)~\cos 2 h(\vec k)\right]
\label{chiralprop2}
\end{eqnarray}
\noindent where $h(\vec k)$ describes the quark
condensate\cite{ijmpe}. Clearly $h(\vec k)\rightarrow 0$ gives
the free massless quark propagator.  In the limit of
vanishing of constituent quark masses the model predicts
absence of chiral symmetry breaking.

We next evaluate the correlation functions of hadronic currents in the vacuum
defined by Eqs. (\ref{vacp})-(\ref{chiralprop2}). It turns out that without any
approximation, the condensate structure uniquely determines the interacting
quark propagator and the mesonic(baryonic) correlators are essentially
squares(cubes) of the propagator. With a Gaussian ansatz for $h(\vec k)$
\cite{ijmpe}, namely $\sin 2h(\vec k)=exp(-R^2 k^2/2)$  we find that the behaviour
of the quark propagator is similar to that obtained by Shuryak and
Verbaarschott\cite{shuqprop}. Further, the mesonic correlation functions were
also qualitatively similar to phenomenological correlation functions in all
channels except for the pseudoscalar(PS) channel. In this channel we did not get
the strong attraction at intermediate ranges seen in the data.
All these results depend very weakly on the choice of the functional
form of $h(\vec k)$.

In view of this outcome, it is obvious that some crucial physics is missing
 from our model, and hence the vacuum structure considered by us has to be
supplemented by additional effects. In our framework, this means that the
quark propagators alone do not determine the hadronic correlators.
This implies that there ought to be explicit contribution arising from
irreducible four point structure of the vacuum. Alternatively we can
describe the irreducible four point vacuum structure as a manifestation of
condensate fluctuations.

In order to proceed further with this idea of condensate fluctuations
we suggest that the normal ordered operators with
respect to the vacuum (Eq. (\ref{vacp})) do not annihilate the
actual ground state of QCD which could be more complicated.
In this case, we have a more general equation

\begin{equation}
T \bar \psi_{\alpha}(\vec x) \psi_{\beta}(\vec x) \bar \psi_{\gamma}(0)
\psi_{\delta}(0) = S_{\beta \gamma}(\vec x) S_{\delta \alpha}(-\vec x)
+ : \bar \psi_{\alpha}(\vec x) \psi_{\beta}(\vec x) \bar \psi_{\gamma}(0)
\psi_{\delta}(0):
\end{equation}
where the : denotes normal ordering with respect to the 
vacuum of Eq.(\ref{vacp}). To include the effect of
fluctuations we may write
\begin{equation}
 : \bar \psi_{\alpha}(\vec x) \psi_{\beta}(\vec x) \bar \psi_{\gamma}(0)
\psi_{\delta}(0): =\Sigma_{\beta \gamma}(\vec x) \Sigma_{\delta \alpha}(-\vec x)
\end{equation}
so that $<vac|\Sigma \Sigma |vac>=0$ but $<\Omega|\Sigma \Sigma |\Omega>
\neq 0$
where $|\Omega>$ is the ``new improved'' QCD ground state including the
condensate fluctuations.

With such a structure for the ground state of QCD the correlator takes
the form

\begin{equation}
R(\vec x)  =   Tr \left[ S(\vec x) \Gamma^{'} S(-\vec x) \Gamma \right]
        + Tr \left[ \Sigma(\vec x) \Gamma^{'} \Sigma(-\vec x) \Gamma \right]
\end{equation}
with the extra term arising from the condensate fluctuations. The
structure for the fluctuating field is arbitrary so far with all possible
Dirac matrix structures (i.e. 1, $\gamma_5$, $\gamma_\mu$, $\gamma_\mu
\gamma_5$,  $\gamma_{\mu \nu}$ ; $\mu \neq \nu$ )

The experimental data demands that we choose the condensate fluctuation
 field such that
it contributes maximally in the PS channel and should not affect
the other channels very much. Such a condition restricts the choice for
the fluctuating field to a structure of the type

\begin{equation}
\Sigma_{\alpha \beta}(\vec x) = \Sigma_{\alpha \beta}^V(\vec x) + \Sigma_{\alpha
  \beta}^S(\vec x) \\
 \quad  = \mu_1^2 \; (\gamma^i \gamma^j)_{\alpha \beta}\;  \epsilon_{ijk}\;
  \phi^k(\vec x) + \mu_2^2 \; \delta_{\alpha \beta}  \; \phi(\vec x)
\end{equation}
where the first term corresponds to vector fluctuations and the second to
scalar.  $\mu_1$ and $\mu_2$ in the above equations are dimensional 
parameters which give the strength of
the fluctuations and $\phi^k(\vec x)$ and $\phi(\vec x)$ are vector and 
scalar fields such that 

\begin{equation}
<\Omega | \phi^k(\vec x) | \Omega> = 0  ; \quad 
<\Omega | \phi(\vec x) | \Omega> = 0
\end{equation}
and
\begin{equation}
<\Omega | \phi^i(\vec x) \phi^j(0) | \Omega> = \delta^{ij} g_V(\vec x)
 ; \quad
<\Omega | \phi(\vec x) \phi(0) | \Omega> = g_S(\vec x)
\end{equation}
This implies an approximation that the ground state of QCD is also a
condensate in the fluctuating fields. The functions $g_{S,V}(x)$ are at this
stage arbitrary. It should be pointed out that the vector
fluctuations contribute only to the PS and nucleon channels, 
while not contributing to the other physical channels. On the other hand the
scalar fluctuations contribute to all channels but most to the vector and
delta channels. It ought to be noted that to be consistent with
phenomenology of correlators, the contribution of vector fluctuations
should be greater than that of scalar fluctuations.

Since we do not expect fluctuations to be important for 
 $\vec x\rightarrow$ 0 and for large $x$ , we want $g(\vec x)$ to vanish
in these two limits.
With these properties in mind we take the function $g(\vec x)$ as

\begin{equation}
g_V(\vec x) = \frac{1}{2 \pi^2 x} \left[\mu_1  K_1(\mu_1 x) -\mu_3 
K_1(\mu_3 x) \right] ; \quad
g_S(\vec x) = \frac{1}{2 \pi^2 x} \left[\mu_2  K_1(\mu_4 x) -\mu_5 
K_1(\mu_6 x) \right]
\end{equation}
so that the small $x$ behaviour of the correlator is the same as expected from
the asymptotic freedom. 
$g_V(\vec x)$ corresponds to taking the fluctuating field condensate function as
difference of two propagators with different masses. Such a structure for the
condensate function arises naturally e.g. in the $\phi^4$ field theory in the
Gaussian effective potential calculations\cite{stev} where $\mu_1$ corresponds to
the lagrangian mass parameter and $\mu_3$ corresponds to the variational
parameter associated with the Gaussian ansatz.
For $g_S(\vec x)$ we had to consider a more general form (with extra
parameters) because the contributions of the propagators and of the
scalar fluctuations are comparable, requiring delicate cancellations.

\begin{figure}
\mbox{\hskip -0.30in}\epsfbox[11 39 600 630]{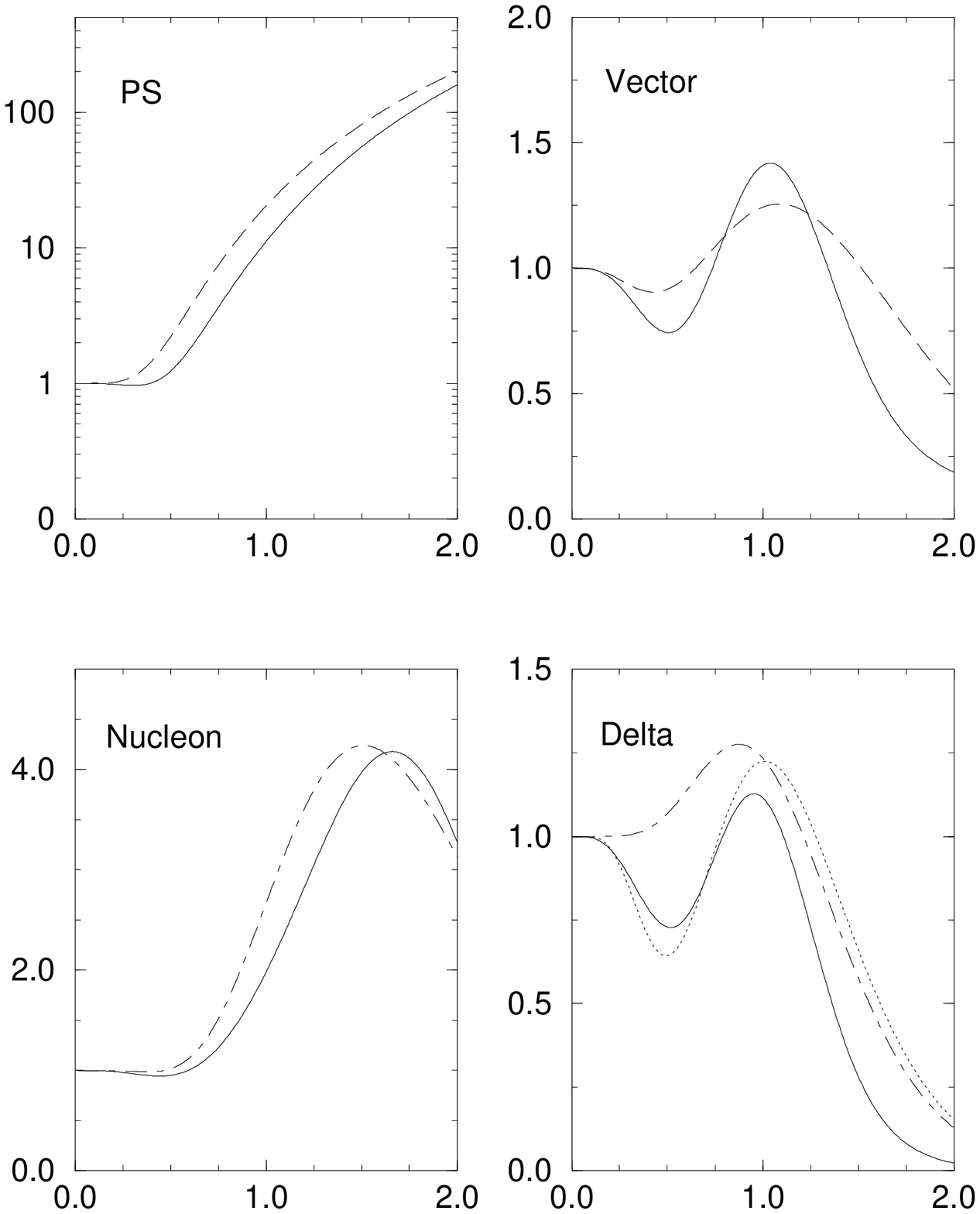}
\caption{The ratio of the hadron correlation functions in QCD
vacuum to the correlation functions for noninteracting massless quarks
$\displaystyle \frac{R(x)}{R_{0}(x)}  $, vs. distance x (in fm). 
Our results are given by the solid curves. The empirical results determined by
dispersion analysis of experimental data in Ref.[3] are shown by
long dashed lines. The results from lattice calculations and instanton model
are denoted by dotted and dot-dashed lines respectively.\label{corfig}}
\end{figure}

 The contributions  of the propagator and the fluctuation fields 
to mesonic and baryonic channels are shown separately in cols. 3 and 4 of 
Table~\ref{table1}.
The parameters of the quark condensate function $h(\vec k)$ and the of
fluctuating fields $g_{S,V}$ are chosen so that our correlation functions are
similar to those obtained from phenomenology.  We have taken $R=0.60 fm$ 
corresponding to  $<\bar \psi \psi>=-(304.45 \; MeV)^3$. 
This value is similar to the value $<\bar \psi \psi>=-(307.4 \; MeV)^3$. 
 considered in Ref.\cite{shumeson,shubar},
though it is higher than the phenomenological value.
Vector fluctuations dominate in the PS channel
beyond $x = 0.5fm$.  This behaviour is reproduced by choosing
$\mu_1 = 164.8$ and $\mu_3 = 69$. In the vector channel the propagator and
fluctuation contributions are of the same order. If we choose $\mu_2 =
245, \mu_4 = 400,  \mu_5 =1280.77, \mu_6 = 518$ 
we get reasonable agreement with
phenomenological curves of correlators. The parameters $\mu_{1...6}$ are
in units of MeV. 
The resulting correlators  are plotted in Fig.~\ref{corfig}.

We fit our results (solid curves of Fig.~\ref{corfig}) to
phenomenologically motivated forms for correlators parametrised in terms of
mass, coupling and threshold of the corresponding particle\cite{neglecor}.
We used the Marquardt method for non-linear least square fit. The method was
very stable and the goodness of fit estimated from $\chi^2$ was reasonable
for small and intermediate $x$, which is the region characterising the 
mass, coupling and threshold of a general correlator\cite{neglecor}. 
Our fitted parameters and those similarly obtained in  lattice\cite{neglecor},
instanton liquid model for QCD vacuum\cite{shumeson,shubar} and 
QCD sum rule\cite{ioffe} calculations are tabulated in Table~\ref{table2}.

As is evident from Fig.~\ref{corfig} and Table~\ref{table2} our model of
the vacuum gives results for the hadronic correlators that are comparable to
those in the instanton model\cite{shumeson,shubar} and lattice
calculations\cite{neglecor}. The contribution of condensate fluctuations is
most important and in some ways is related to the ``hidden contribution''
discussed by Shuryak\cite{shuqprop,shumeson}.To summarise
 we have quite clearly shown that to be
consistent with data, QCD vacuum must not only have quark condensates
but must of necessity also have condensate fluctuations.
\acknowledgments

VS and HM wish to acknowledge discussions with A. Mishra, 
and S. P. Misra. VS would like to thank Dr.(Mrs.) R. Suhasini for
discussions on nonlinear curve fitting. 
HM  thanks Theory Division of Physical Research Laboratory Ahmedabad for
a visit and also
acknowledges to the Council of Scientific and Industrial Research, 
Government of India for the research associateship 9/679(3)/95-EMR-I.

\def\feynman{ R.P. Feynman, Nucl. Phys. B188 (1981) 479.}

\def\sur{ E.V. Shuryak, {\it The QCD vacuum,
 hadrons and the superdense matter}, (World Scientific, 
Singapore, 1988).}

\def\surcor{E.V. Shuryak, Rev. Mod. Phys. 65 (1993) 1 .} 

\def \neglecor{M.-C. Chu, J. M. Grandy, S. Huang and 
J. W. Negele, Phys. Rev. D48 (1993) 3340 ;
ibid, Phys. Rev. D49 (1994) 6039 .}

\def \ijmpe{ A. Mishra, H. Mishra, S.P. Misra, P.K. Panda 
and Varun Sheel, Int. J. Mod. Phys. E5 (1996) 93.}

\def \shuqprop{E. V. Shuryak and J. J. M. Verbaarschot,
Nucl. Phys. B410 (1993) 37.}

\def\stev { P.M. Stevenson, Phys. Rev. D32 (1985) 1389 ; 
H. Mishra and A.R. Panda, J. Phys. G 
(Part. and Nucl. Phys.)18 (1992) 1301 .}

\def \shumeson{E. V. Shuryak and J. J. M. Verbaarschot,
Nucl. Phys. B410 (1993) 55.}

\def \shubar{T. Sch$\ddot{a}$fer, E. V. Shuryak and J. J. M.
Verbaarschot, Nucl. Phys. B412 (1994) 143.}

\def \adler{S. L. Adler and A.C. Davis, Nucl. Phys. B244 (1984) 469;
R. Alkofer and P. A. Amundsen, Nucl. Phys. B306 (1988) 305 ; M. G. Mitchard,
A. C. Davis and A. J. Macfarlane, Nucl. Phys. B325 (1989) 470 .}

\def \ioffe{B.L. Ioffe, Nucl. Phys. B188 (1981) 317 ; V.B. Beleyaev and
B.L. Ioffe, Zh. Eksp. Teor. Fiz. 83 (1982) 976 
 [Sov. Phys. JETP 56 (1982) 547 ] .}

\pagebreak[4]
\begin{table}
\caption{Meson currents and correlation functions \label{table1}}
\begin{tabular}{llll}
CHANNEL & CURRENT & \multicolumn{2}{c}{CORRELATION FUNCTIONS
 $\displaystyle \left[ \; \frac{R(x)}{R_{0}(x)} \; \right]$ }\\
 & & Without fluctuations\tablenotemark[1]&Fluctuation contribution\\   
 & & &(Vector($ F^V$) and Scalar ($ F^S$)) \\
 & & & \\ 
\tableline
Vector & $\bar u \gamma_{\mu} d$  
& $[F(x)]^2 + \frac{\pi}{4} \frac{x^6}{R^6} e^{-x^2/R^2}$ 
&$ F^V= 0$  \\
 & & & $ F^S= 8 \pi^4 x^6 g_S(2x)$ \\ 
\hline  
Pseudoscalar & $\bar u \gamma_{5} d$  
&  $[F(x)]^2 + \frac{\pi}{8} \frac{x^6}{R^6} e^{-x^2/R^2}$
&$ F^V= -48\pi^4 x^6 g_V(2x)$  \\
 & & & $ F^S= 4 \pi^4 x^6 g_S(2x)$ \\ 
\hline  
Nucleon  & $ \epsilon_{abc} \left[\tilde u^a(x) C \gamma_\mu u^b(x)
\right] \gamma^\mu \gamma_5 d^c(x) $
& $ [F(x)]^3 + \frac{\pi}{16} \frac{x^6}{R^6}  e^{-x^2/R^2} F(x) $
 &$ F^V= -4\pi^4 x^6 g_V(2x) F(x)$  \\
 & & & $ F^S= 2 \pi^4 x^6 g_S(2x) F(x)$ \\ 
\hline  
Delta & $
\epsilon_{abc} \left[\tilde u^a(x) C 
\gamma_\mu u^b(x) \right]u^c(x) $
& $ [F(x)]^3 + \frac{\pi}{4} \frac{x^6}{R^6}  e^{-x^2/R^2} F(x) $
&$ F^V= 0$  \\
 & & & $ F^S= 8 \pi^4 x^6 g_S(2x) F(x)$ \\ 
\end{tabular}
\tablenotetext[1]{ $F(x)=\left[1+\frac{1}{2} x^2 I(x) \right] $
where $I(x) = \int_{0}^{\infty } \left( \cos k x - \frac{\sin k
x}{kx} \right) \frac{k e^{-R^2 k^2}}{1+(1-e^{-R^2 k^2})^{1/2}} dk $}
\end{table}

\begin{table}
\caption{Fitted Parameters \label{table2}}
\begin{tabular}{lllll}
CHANNEL & SOURCE & M (GeV) & $\lambda$ & $\sqrt{s_0}$(GeV) \\
\tableline
Vector & Ours & 0.78 $\pm$ 0.005 & (0.42 $\pm$ 0.041 GeV)$^2$
& 2.07 $\pm$ 0.02 \\
 & Lattice & 0.72 $\pm$ 0.06 & (0.41 $\pm$ 0.02 GeV)$^2$
& 1.62 $\pm$ 0.23 \\
 & Instanton & 0.95 $\pm$ 0.10 & (0.39 $\pm$ 0.02 GeV)$^2$
& 1.50 $\pm$ 0.10 \\
 & Phenomenology & 0.78 & (0.409 $\pm$ 0.005 GeV)$^2$
& 1.59 $\pm$ 0.02 \\
\hline  
Pseudoscalar & Ours & 0.137 $\pm$ 0.0001 & (0.475 $\pm$ 0.015 GeV)$^2$
& 2.12 $\pm$ 0.083 \\
 & Lattice & 0.156 $\pm$ 0.01 & (0.44 $\pm$ 0.01 GeV)$^2$
& $\; \; \;<$ 1.0  \\
 & Instanton & 0.142 $\pm$ 0.014 & (0.51 $\pm$ 0.02 GeV)$^2$
& 1.36 $\pm$ 0.10 \\
 & Phenomenology & 0.138 & (0.480 GeV)$^2$
& 1.30 $\pm$ 0.10 \\
\hline  
Nucleon & Ours & 0.87 $\pm$ 0.005 & (0.286 $\pm$ 0.041 GeV)$^3$
& 1.91 $\pm$ 0.02 \\
 & Lattice & 0.95 $\pm$ 0.05 & (0.293 $\pm$ 0.015 GeV)$^3$
& $\; \; \;<$ 1.4 \\
 & Instanton & 0.96 $\pm$ 0.03 & (0.317 $\pm$ 0.004 GeV)$^3$
& 1.92 $\pm$ 0.05 \\
& Sum rule & 1.02 $\pm$ 0.12 & (0.337 $\pm$ 0.0.014 GeV)$^3$
& 1.5\\
 & Phenomenology & 0.939 & $ \; \; \; ?$
& 1.44 $\pm$ 0.04 \\
\hline  
Delta & Ours & 1.52 $\pm$ 0.003 & (0.341 $\pm$ 0.041 GeV)$^3$
& 3.10 $\pm$ 0.008 \\
 & Lattice & 1.43 $\pm$ 0.08 & (0.326 $\pm$ 0.020 GeV)$^3$
& 3.21 $\pm$ 0.34 \\
 & Instanton & 1.44 $\pm$ 0.07 & (0.321 $\pm$ 0.016 GeV)$^3$
& 1.96 $\pm$ 0.10 \\
 & Sum rule & 1.37 $\pm$ 0.12 & (0.337 $\pm$ 0.014 GeV)$^3$
& 2.1  \\
 & Phenomenology & 1.232  & $ \: \; \;?$
& 1.96 $\pm$ 0.10 \\
\end{tabular}
\end{table}

\end{document}